\documentclass[a4paper,11pt]{article}
\usepackage{graphicx}

\begin{document}

\title{Discrete breathers in thermal equilibrium:
distributions and energy gaps }

\author{M. Eleftheriou$^{1,2}$  and S. Flach$^3$\\
$^1$  Department of Physics, University of Crete P. O. Box 2208,\\ 
71003 Heraklion, Crete, Greece \\
$^2$ I. N. F. M. Unita di Firenze, Via G. Sansone 1, 50019 Sesto F.no, Italy\\
$^3$ Max Planck Institute for the Physics of Complex Systems, \\
N\"othnitzer Str. 38, D-01187 Dresden, Germany
}
\maketitle

\begin{abstract}
We study a discrete two-dimensional nonlinear system that 
allows for discrete breather solutions. We perform a spectral
analysis of the lattice dynamics at thermal equilibrium and
use a cooling technique to measure the amount of breathers
at thermal equilibrium. Our results confirm the existence of
an energy threshold for discrete breathers. The cooling method
provides with a novel computational technique
of measuring and analyzing discrete breather distribution properties
in thermal equilibrium.
\end{abstract}

\section{Introduction}

Nonlinear discrete systems support discrete breathers (DBs). 
These time-periodic and spatially localized 
solutions are the result of the interplay between nonlinearity 
and discreteness \cite{DB-REVIEWS}. Many studies of DBs 
have been successfully launched,
on such topics as  
rigorous existence proofs, dynamical and structural 
stability and computational methods of obtaining DBs 
in classical models as well as their quantum aspects. 
In addition DBs have been detected and studied experimentally
in such different systems as interacting Josephson junction systems
\cite{binder00a},
coupled nonlinear optical waveguides \cite{eisenberg98}, 
lattice vibrations in
crystals \cite{swanson99}, antiferromagnetic structures \cite{schwarz99},
micromechanical cantilever arrays \cite{sato03}, Bose-Einstein
condensates loaded on optical lattices \cite{BEC}, layered high-$T_c$
superconductors \cite{hightc}. Their existence is also predicted
to exist in the dynamics of dusty plasma crystals \cite{plasma}.

Among several intriguing unresolved questions concerning DBs,
a central issue is the contribution of DBs to the dynamics
of systems at thermal equilibrium. Indeed breather-like excitations
have been observed in a variety of different models at finite
temperatures \cite{equilibrium_old}. 
For some special models with additional conservation laws
semianalytical statements about the contribution of
DBs to thermal equilibrium have been derived \cite{DNLS}.
The question is then whether we can identify the
contribution of DBs to various equilibrium and relaxation properties
(like e.g. the charge trapping in DNA \cite{kalosakas}).
Since DBs are dynamical excitations, their contribution will be
observable mainly in time-dependent (or frequency-dependent) correlation
functions, while static correlation functions, e.g. specific heat,
only probe the available energy landscape, and are not suitable for
detecting dynamical correlations.
A good way to proceed is to use a specific property of DBs
and trace its contribution to correlation functions.
Another question is whether we can design computational methods
to separate the breather excitations at a given time from the
rest of the excitations in the lattice. This would allow us
to perform systematic studies of distribution properties of
DBs in a given lattice at a given temperature.

Concerning the first point from above, we know that
in general nonlinear systems in two and three dimensions support breather 
solutions that have a positive lower energy threshold \cite{FKM}. 
This is a very important property that can be of help 
in the detection of DBs in 
experiments. In fact for specific one-dimensional lattices this
property holds as well. However the search for traces of these thresholds
in correlation functions for one-dimensional lattices 
at thermal equilibrium turned out to be
very complicated \cite{EFT}. There are two reasons for that.
First, DBs in one-dimensional lattices act as strong scatterers
of plane waves \cite{scattering}. 
Consequently radiation can be efficiently trapped between
DBs, and contribute to a strong interaction between DBs. It is also
hard to find a way for letting the radiation out of the system
(see \cite{1drelaxation}).
Secondly frequency-dependent correlation functions probe
gaps in frequency space. Although the existence of energy thresholds
for breathers leads also to frequency thresholds, the values of
these frequency thresholds may become too small to be easily detected
\cite{EFT}.

In this work we examine a two-dimensional lattice.
First fingerprints of the DB energy threshold in two-dimensional
lattices have been reported in \cite{P2} and also in
\cite{BV}. A comprehensive study of
DBs in thermal equilibrium and their influence on
various relaxation mechanisms was provided by Ivanchenko et al
\cite{If}. 
We attempt to go sufficiently beyond these studies.
Here we present
not only the frequency dependence of correlation functions, but
also their temperature (or energy) dependence. We then proceed
to apply a cooling technique at the boundaries of our system to
efficiently get rid of extended excitations in the lattice.
The remaining localized breather-like excitations can then
be easily analyzed. We observe the existence of energy thresholds
and provide with novel distribution functions for DB energies.

\section{Gap determination in the Hamiltonian system}
We study a two-dimensional
quadratic lattice with one degree of freedom 
per site. The equation of motion for the particle at site $(i,j)$ in a 
lattice of size $N\times N$ with free ends is given by
\begin{equation}
\ddot u_{i,j}=k(u_{i+1,j}+u_{i-1,j}-2u_{i,j}) +k(u_{i,j+1}+u_{i,j-1}-2u_{i,j})
-u_{i,j}-u_{i,j}^3,
\label{2-1}
\end{equation}
where $u_{i,j}$ is the displacement of the particle at site ${i,j}$. The masses
 of the particles are equal to unity and the interaction between them is 
harmonic with strength $k$ while the on-site potential is of 
hard-$\phi^{4}$ type.
The corresponding Hamiltonian from which (\ref{2-1}) is derived, is given by
\begin{equation}
H = \sum_{i,j} h_{i,j}\;,\; h_{i,j}=\frac{1}{2}\dot{u}^2_{i,j}
+ \frac{1}{2}u^2_{i,j} + \frac{1}{4}u^4_{i,j}
+\frac{k}{4}\sum_{NN}(u_{i,j}-u_{l,m})^2\;.
\label{2-2} 
\end{equation}
Here $h_{i,j}$ is the discrete on-site 
energy density which will be used later,
and $NN$ stands for nearest neighbors and implies $(l-i)^2+(m-j)^2=1$.
For small amplitudes the linearized equation (\ref{2-1}) yields
plane waves $u_{l,m}(t) \sim {\rm e}^{i(\omega_q t - q_xl - q_y m)}$
with the phonon dispersion relation
\begin{equation}
\omega_{\vec{q}}^2 = 1+4k\left( \sin^2(\frac{q_x}{2}) + \sin^2(\frac{q_y}{2})
\right) 
\label{2-2a}
\end{equation}
where $\vec{q}\equiv (q_x,q_y)$ is the wave vector of the reciprocal lattice.

\begin{figure}
\includegraphics[angle=0, width=1\textwidth]{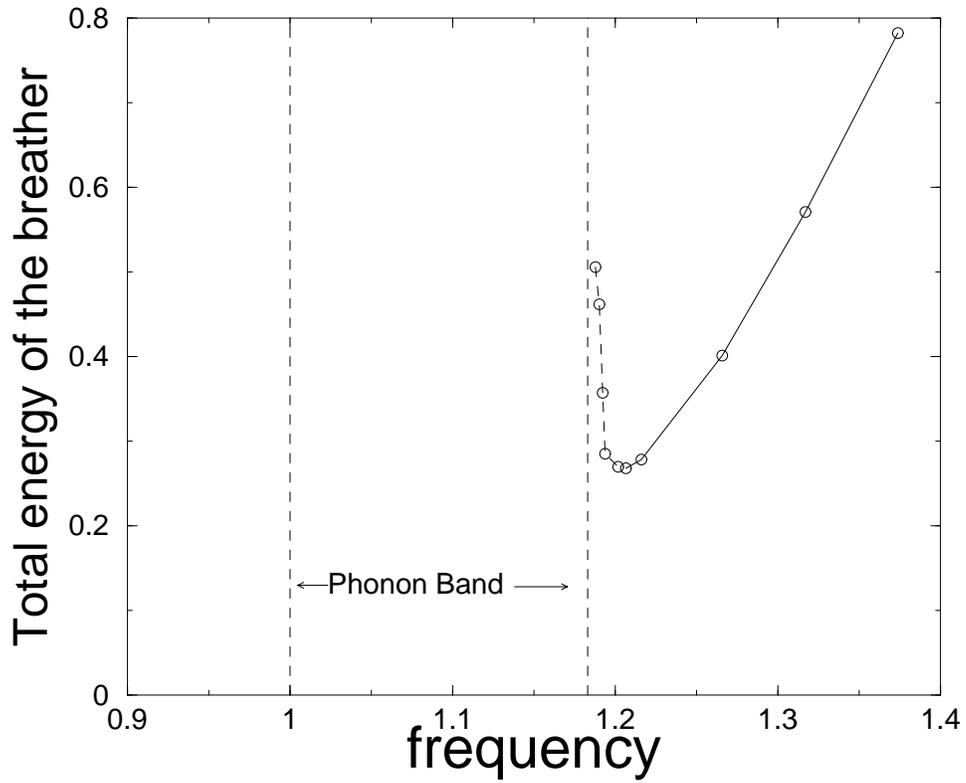}
\caption{Energy of the breather solution for coupling $k=0.05$ as a function 
of its frequency (circles). 
Lines connecting circles are guides to the eye.
The dashed vertical lines indicate the boundaries 
of the phonon band.}
\label{fig0}
\end{figure}

\begin{figure}
\includegraphics[angle=0, width=1\textwidth]{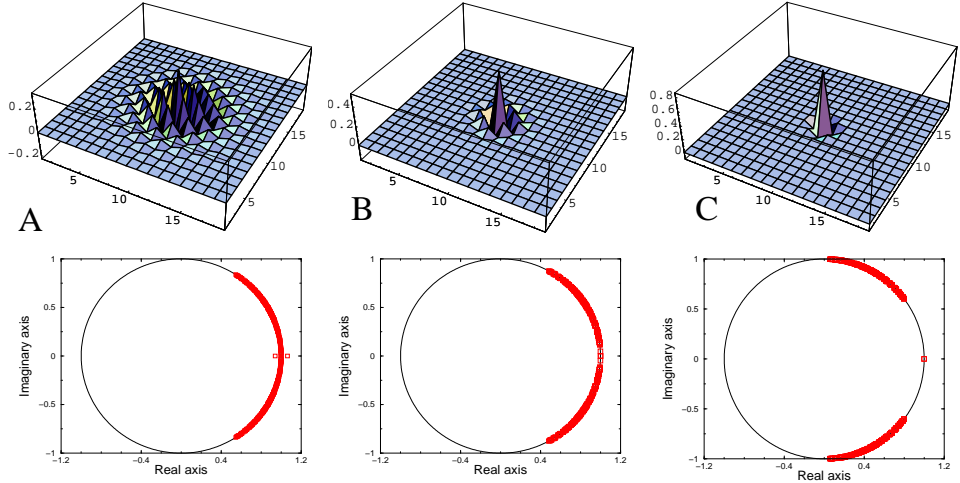}
\caption{Displacements $u_{lm}(t=0)$ of breather solutions
at initial time with zero velocities $\dot{u}_{lm}(t=0)=0$.
Below each profile the eigenvalues of the Floquet
matrix of linearized perturbations are shown in the complex plane (squares)
together with the reference unit circle.
A: $\Omega_b=1.188$; B: $\Omega_b=1.207$; C: $\Omega_b=1.319$. 
}
\label{figbr3d}
\end{figure}

We obtain the energy threshold 
for breathers using the Newton 
algorithm and the anticontinuous limit \cite{DB-REVIEWS}. We 
compute 
breather solutions for different frequencies for 
coupling $k=0.05$. The phonon band
of small amplitude plane waves is located 
between the frequencies $\omega_{\vec{q}=(0,0)}\equiv \omega_0=1$ and 
$\omega_{\vec{q}=(\pi,\pi)}\equiv \omega_{\pi}=1.183$. 
We use a 
lattice of $19\times19$ sites. In Fig. \ref{fig0} we plot the 
total energy of the breather solution as a function of its frequency $\Omega_b$. 
Additionally we check the stability of
the obtained solutions using the standard 
Floquet analysis \cite{DB-REVIEWS}. 
We observe in Fig. \ref{fig0} that the curve of the 
total energy of the breather as a function of frequency consists of 
two branches, separated by a minimum at frequency $\Omega_{b,th}$. 
In the left branch (dashed line) DBs 
are unstable according to the Floquet analysis (see Fig.\ref{figbr3d}A)
while in the right branch (solid line) 
DBs are linearly stable solutions (Fig.\ref{figbr3d}C). 
The breather with frequency $\Omega_{b,th}=1.207$ is 
separating the frequency region $\omega_{\pi} < \Omega_b < \Omega_{b,th}$ 
where breathers are unstable and the region 
$\Omega_{b,th} < \Omega_b$
where breathers are linearly stable according to our computations. This 
specific breather solution has a total energy $E_{b,th}=0.268$.
$60 \%$ of the energy of this DB 
is located on the  central site, i.e. $h_{central}=0.167$.
In other words, the minimum energy breather is still a rather
discrete object, involving essentially only a few lattice sites
in its dynamics (Fig.\ref{figbr3d}B) (see also \cite{FKM},\cite{kalosakas2}).

In terms of energy the threshold value $E_{b,th}$ provides with
a rigorous lower bound - no breathers exist with energies less than
$E_{b,th}$.
Consequently we may expect this feature to be detectable
in the temperature dependence of equilibrium correlation functions.
The corresponding order of magnitude of the average energy per site
is expected to be in the region of values $0.05 - 0.2$.
While the upper value is simply close to $E_{b,th}$, the lower one
can be obtained by observing that 
essentially one central cite and four neighboring sites
are important for the dynamics of stable breathers, providing with
a lower estimated value of $E_{b,th}/5$.
At the same time the frequency gap $\omega_{\pi} < \Omega_b < \Omega_{b,th}$
provides with a less rigorous bound. We can only speculate that
breathers with frequencies belonging to this gap region are less probable
to be excited, because they are linearly unstable. Nevertheless they
do exist, and in thermal equilibrium the system may be excited
close to these solutions for short times.

\section{Thermal equilibrium}

\begin{figure}
\includegraphics[angle=0, width=1\textwidth]{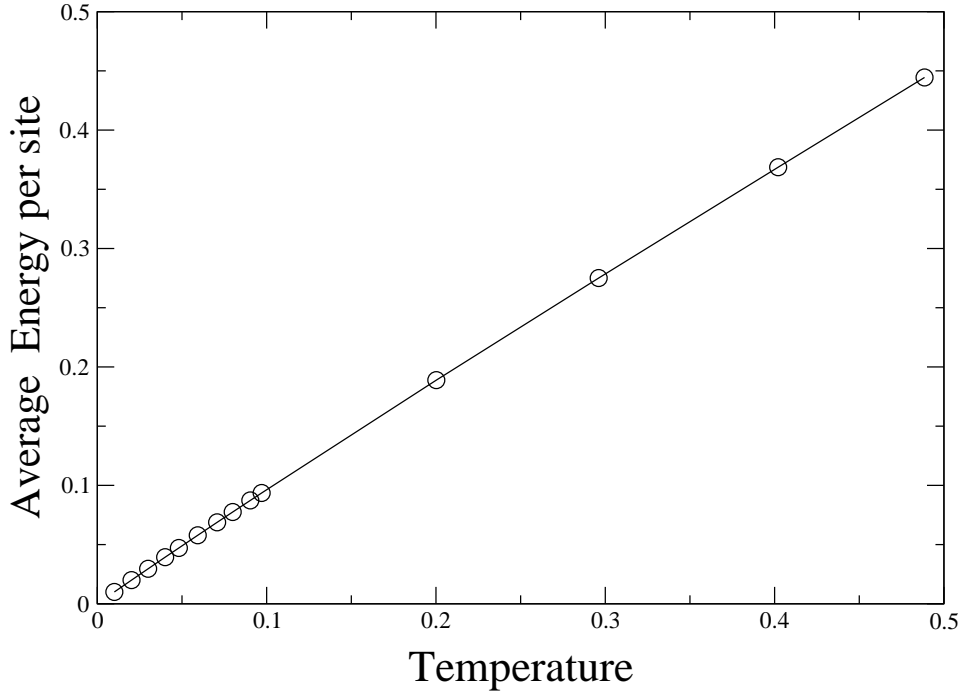}
\caption{Dependence of the average energy per site $\langle h_{l,m} \rangle$
on the temperature in the parameter range of interest.}
\label{E-T}
\end{figure}

We thermalize the lattice using as a first method
Langevin equations of motion .
We add to the right hand side of equations (\ref{2-1})
damping terms $-\gamma \dot{u}_{i,j}$ and a Gaussian white
noise force $\xi(t)$. The friction is chosen to be $\gamma=0.01$.
The
Gaussian white noise
$\xi$ is characterized by the standard correlation function
$\langle \xi(t) \xi(t') \rangle = 2\gamma k_B T \delta (t-t')$ 
where $k_B=1$ is the dimensionless Boltzmann constant
and $T$ is the temperature.
We simulate the Langevin equations until the averaged kinetic energy
per particle is close to the desired value $T/2$. We then
switch off the friction and Gaussian white noise terms, and
continue integrating the Hamiltonian equations (\ref{2-1}).
Time is set to $t=0$ at this moment of switching from Langevin
to Hamiltonian evolution.
We then measure the time and ensemble averaged kinetic energy
and reobtain the corresponding value of $T$, together with computing
the time and ensemble averaged energy per site.
The temperature values are very close to
the corresponding averaged total energies per site $\langle h_{l,m} \rangle$
as shown in Fig.\ref{E-T}.

We compute the time dependent 
displacement-displacement correlation function,
\begin{equation}
S(t)=\frac{1}{S_0}\frac{1}{N^{2}}[\frac{1}{t_{fin}-t_{in}}
\sum_{i=1}^{N}\sum_{j=1}^{N}
\int_{t_{in}}^{t_{fin}}\langle u_{i,j}(t+t')u_{i,j}(t')\rangle dt'],
\label{1eq}
\end{equation}
where the brackets denote an ensemble average. In this study the ensemble 
averaging is obtained by averaging the results for
10 different realizations or runs. The parameters are $t_{in}=8000$ and 
$t_{fin}=8500$. 
The Fourier transform of $S(t)$ is:
\begin{equation}
S(\omega_{m})=\Delta t 
\sum_{j=0}^{\frac{t_{fin}}{\Delta t}-1} 
e^{-\frac{2\pi i}{t_{fin}} m \Delta t j} S(\Delta t j),
\end{equation}
where 
$\omega_{m}=\frac{2 \pi m}{t_{fin}}$, 
$m=0,1,......,\frac{1}{2}\frac{t_{fin}}{\Delta t}-1$. 
Finally the power spectrum of the correlation function $S(t)$ is 
given by
\begin{equation}
P(\omega)=(Re(S(\omega)))^2+(Im(S(\omega)))^2,
\end{equation}
where $Re(S(\omega))$ and $Im(S(\omega))$ denote the real and imaginary part 
respectively. The parameter $S_0$ is chosen such that
$S(t=0)=1$ implying that the frequency integral 
over $S(\omega)$ is constant.

\begin{figure}
\includegraphics[angle=0, width=1\textwidth]{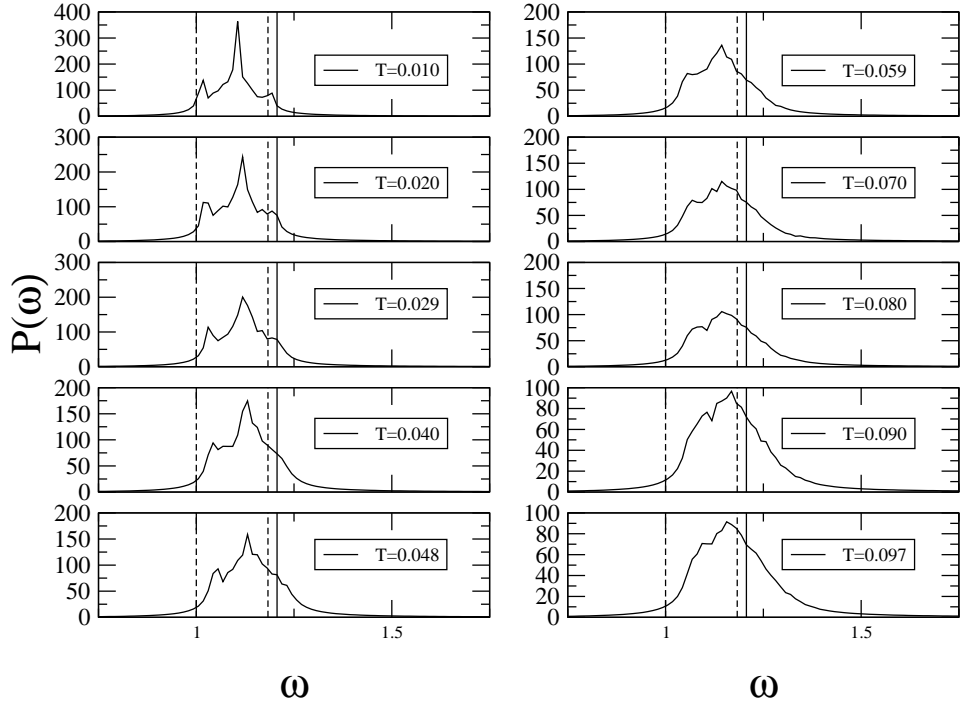}
\caption{Power spectra $P(\omega)$ as a function of frequency  for various 
values of temperature. The coupling between the sites is $k=0.05$. The values 
of temperature are indicated in the labels in each subgraph. The two 
vertical dashed lines mark the band edge values $\omega_0$ and $\omega_{\pi}$
of the phonon band while the vertical solid line 
marks $\Omega_{b,th}=1.207$.
}
\label{fig1}
\end{figure}
In Fig.\ref{fig1} we plot the power spectrum $P(\omega)$ for various 
temperatures. For low temperature values the power spectrum 
shows that the modes with frequencies inside the phonon band are excited. 
As the temperature increases the power spectrum is shifted 
to larger frequencies due to the hard anharmonicity. 
Additionally 
we observe that for temperatures $T \approx 0.04$ and higher the spectrum 
exhibits a different behavior in the frequency 
gap region, viz. shows a decrement near 
the gap.
\begin{figure}[!h]
\includegraphics[angle=0, width=1\textwidth]{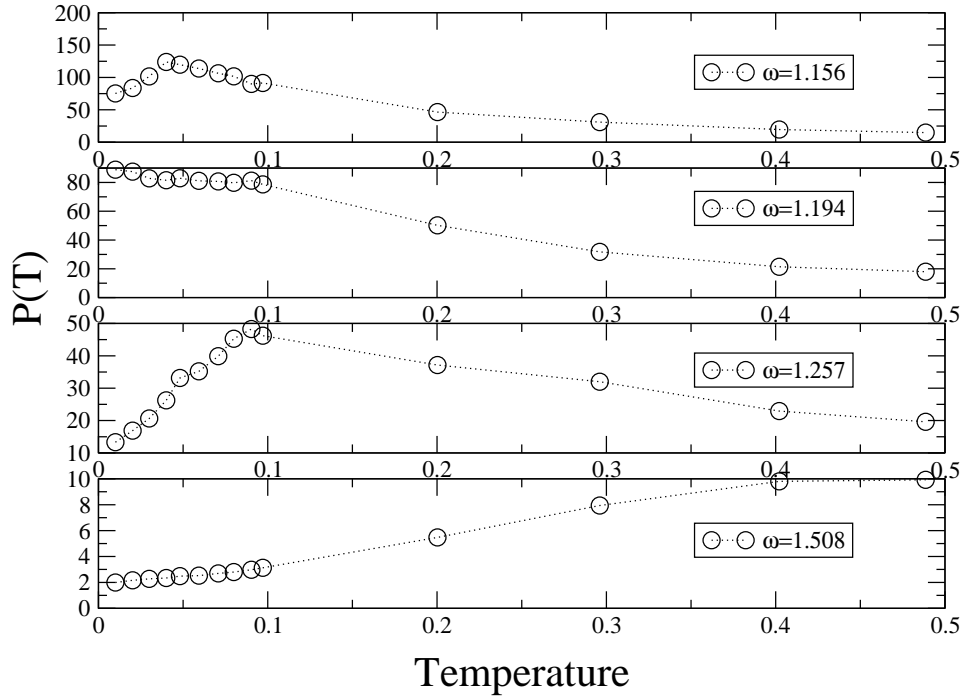}
\caption{Power spectra $P(T)$ as a function of temperature  for $4$ different 
frequencies. From top to the bottom: frequency inside the phonon band, 
$\omega=1.156$; frequency inside the 
frequency gap where breathers 
are unstable $\omega=1.194$; 
frequency  lies outside the two previous regions 
$\omega=1.257$; largest frequency value  
$\omega=1.508$.}
\label{fig3}
\end{figure}
Let us examine the power spectra in detail.
In Fig. \ref{fig3} we plot the value of the power spectrum $P(T)$ as a function 
of temperature for four specific frequencies. The top panel of 
Fig. \ref{fig3} is plotted for a frequency inside 
the phonon band, the second panel is plotted for a frequency that belongs to 
the breather frequency gap, while the lower 
two panels are plotted for frequencies that are located
outside the phonon band and the frequency gap region. 
In the top panel of the Fig. \ref{fig3} we observe a maximum 
as the temperature increases due to the shift of the peak of the phonon 
spectrum with temperature in Fig.\ref{fig1}. 
This peak is located at $\omega=1.156$ for low temperatures.
With increasing  
temperature the peak passes through the specific frequency that we 
chose in the top panel of Fig.\ref{fig3}. 
The characteristic temperature value of this peak will depend
on the chosen value of the probe frequency inside the phonon band.

In the second panel in Fig.\ref{fig3}, where the 
chosen frequency belongs to the breather frequency gap,
 a crossover is observed  at 
$T \approx 0.1$. This crossover is even more evident
in the third panel. In this case the 
frequency lies outside the gap region but, at the same time, is very close 
to it. A crossover is observed again at $T \approx 0.1$, a value
that has the same order of magnitude as the energy threshold value 
$E_{b,th}$ that we 
found in Section 2. 
Finally in the bottom panel 
in Fig.\ref{fig3} a trace of this threshold 
is still observed, although the probe frequency is located far from the 
phonon band and breather frequency gap values, and the power spectrum
$P(\omega)$ has only small tails there.

Thus we do observe a characteristic crossover feature in the 
power spectra, but not as initially expected upon varying the frequency
at a fixed temperature. Instead we observe the crossover for a fixed
probe frequency by varying the temperature. The absence of a clearly
observable frequency gap may be due to the fact that the frequency 
gap itself is
rather narrow, that unstable breathers may still be excited and
persist for some time with some probability, and that anharmonic
extended waves contribute as well.
An interesting question is why the observed crossover upon varying the
temperature is showing a maximum or similar cusp for frequencies
close to the frequency gap, and an (although bearly visible)
opposite behavior at larger probe frequencies (as seen in the bottom
panel in Fig.\ref{fig3}). A possible answer is that for low temperatures
only occasionally breathers are excited, and the shift of frequency
contributions to higher values with increasing temperature is as well
caused by simple hard anharmonicity effects. However once  the
temperature reaches the threshold
value for breathers, breathers with larger energies are easily
excited as well. This will cause a depletion of the power spectra at
lower frequencies, at the expense of the increase at higher frequencies,
just as observed in Fig.\ref{fig3}. 
Still the above argumentation shows that it is hard to separate
the contribution of anharmonic phonons from that of breathers.
We clearly are in need of a technique which does this separation.
The next chapter will provide with a solution to this problem.

\section{How to measure breathers in thermal equilibrium}

As already discussed in the introduction, DBs act as strong
point-like scatterers of plane waves \cite{scattering}. 
In one-dimensional
lattices this circumstance makes it hard to let the delocalized
radiative part of excitations out of the system \cite{If}. 
However in two-dimensional
(and even more efficiently in three-dimensional) systems
point-like scatterers will not hinder plane waves from moving
around such an obstacle. Consequently we may attach dissipative
boundaries to our lattice, and expect the radiation to disappear
from the system during a reasonable short time, leaving the immobile
localized excitations behind.
Such cooling techniques have been used for
studying breather relaxations \cite{1drelaxation},\cite{BV},\cite{If}. 
Here we are only interested
in letting the radiation out and being left with DBs. 

We study an ensemble of $10$ different realizations that correspond to the 
same initial energy per site $E_{0}$. 
We thermalize the system in another way here.
We start the system with a given initial 
energy $E_{0}$, i.e. we integrate the system for 1000 time units using 
initially random displacements for the particles.
After that dissipative boundaries are switched on for a given
transient of time. This is done here by adding a friction term
to the boundary sites of a $20 \times 20$ lattice
with friction constant $0.1$ for some cooling time $T_{cool}$. 
While much more sophisticated
ways of reflectionless dissipative boundaries can be implemented
\cite{kladko},
this simple method suffices for the results presented below.
 
\begin{figure}[!h]
\includegraphics[angle=0, width=1\textwidth]{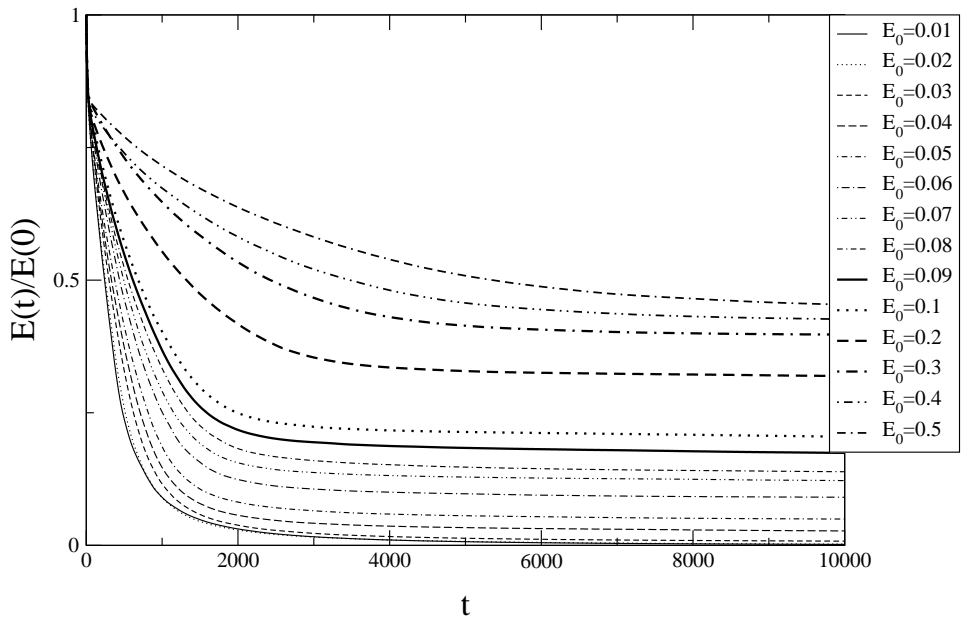}
\caption{The time dependence of the normalized lattice energy per site
in the presence of a dissipative boundary for different
values of the initial energy per site as indicated in the figures.
 }
\label{figdiss1}
\end{figure}

\begin{figure}[!h]
\includegraphics[angle=0, width=1\textwidth]{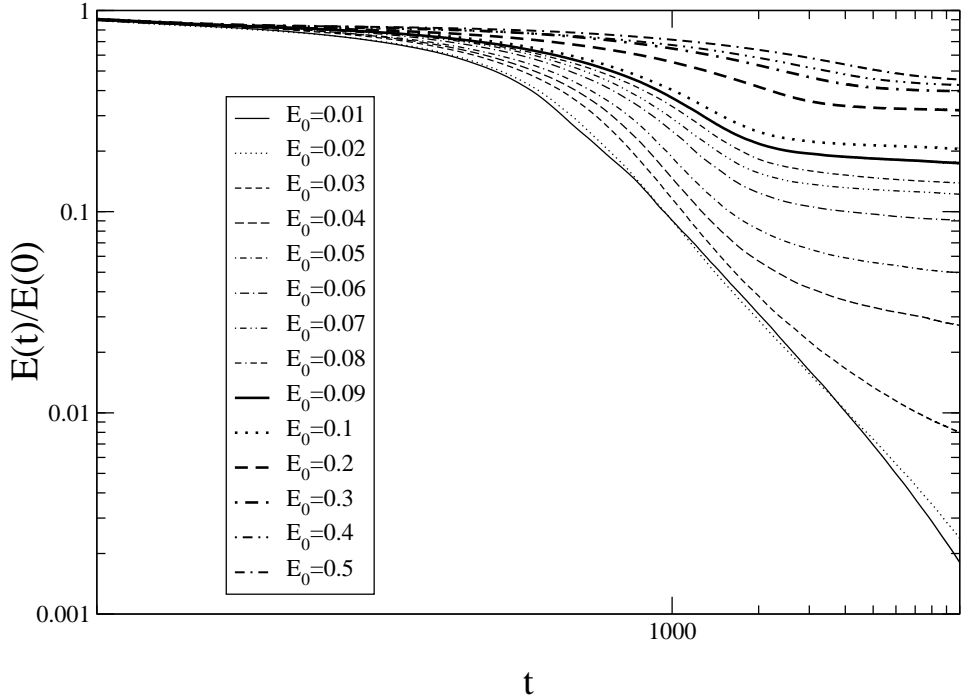}
\caption{Same as in Fig.\ref{figdiss1} but in logarithmic units. }
\label{figdiss2}
\end{figure}

The effect of the dissipative boundary is shown in
Figs.\ref{figdiss1},\ref{figdiss2}
where the energy per site remaining in the lattice is plotted
as a function of time, both in normal and in logarithmic units.
Note that the energy values are normalized to their corresponding
number at time $t=0$ at which the dissipative boundary is switched on.
\vspace*{1cm}

We clearly observe that after some transient behavior the
lattice energy density changes rather slowly, indicating the
expected outcome - delocalized excitations left the system, leaving
the localized ones behind. 
Both Fig.\ref{figdiss1} and especially Fig.\ref{figdiss2} show
that the characteristic waiting time increases with increasing
initial energy density from roughly $T_{cool}=2000$ up to
$T_{cool}=10000$. 
This can be observed from the shift of the inflection points
in Fig.\ref{figdiss2}.
In the following we will use
a cooling 
time $T_{cool}=10^{4}$, but we will also compare with shorter
cooling times. 

\begin{figure}
\includegraphics[angle=0, width=1\textwidth]{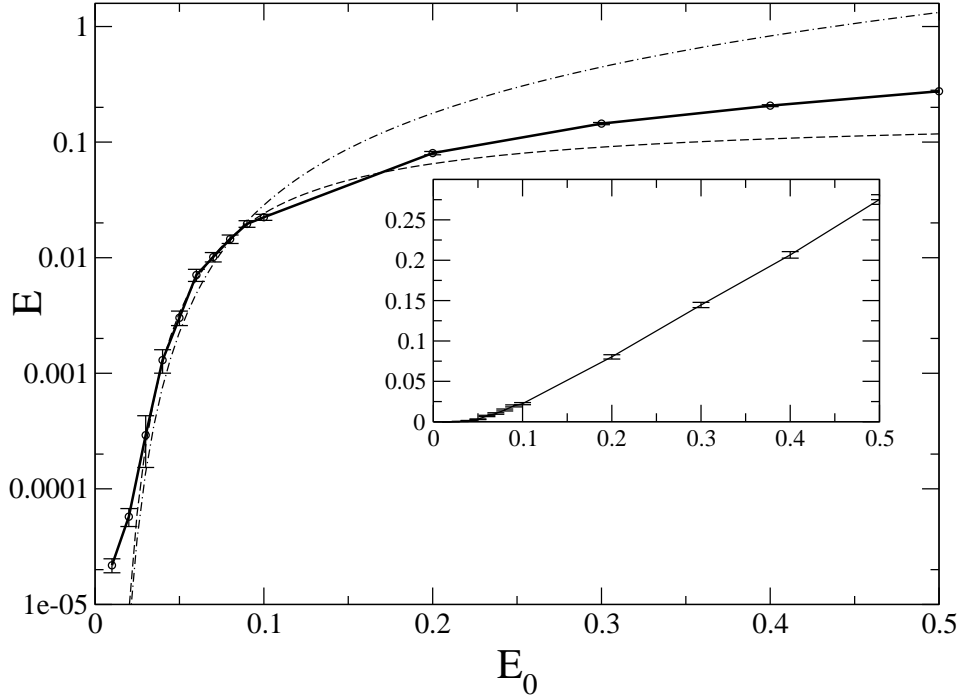}
\caption{Graph of the average energy per site and per number of realizations 
$E$ as a function of the initial energy $E_{0}$ after cooling. The error 
bars are the standard deviations of the mean for the statistical ensemble.
Note the logarithmic scale of the y-axis.
Solid line -guide to the eye.
Dashed line - fit using (\ref{fit1}). Dashed-dotted line - fit
using (\ref{fit2}) (for parameters see text).
Inset: Same but with linear y-axis scaling to observe the
crossover around $E_0=0.05$. 
}
\label{fig4}
\end{figure}

In Fig. \ref{fig4} we plot the total energy that remains in 
the lattice (excluding the sites that belong to the boundaries) per site and 
per number of realizations $E$ as a function of initial energy $E_{0}$.  
Note the log scale for the y-axis. For clarity the inset is for
the same data but with linear axis scaling.
We observe a crossover for $E_{0}\sim 0.05$. 
We fit the curve of $E(E_{0})$ first using
\begin{equation}
E=A \tilde{E}_{b,th} e^{-\frac{\tilde{E}_{b,th}}{E_{0}}}
\label{fit1}
\end{equation} 
with parameters A and $\tilde{E}_{b,th}$. This 
holds assuming that only breathers with $\tilde{E}_{b,th}$ will
be excited. Then
the probability to form a DB is  $e^{-\frac{\tilde{E}_{b,th}}{E_{0}}}$ 
while its contribution to 
an energy distribution will be 
$\tilde{E}_{b,th} e^{-\frac{\tilde{E}_{b,th}}{E_{0}}}$. 
Depending on the energy range we use for fitting we obtain
$\tilde{E}_{b,th}=0.17$ ($E_0 < 0.06$) or $\tilde{E}_{b,th}=0.233$
($E_0 < 0.1$). We were not able to fit data at larger energies.
The obtained 
values 0.17-0.233 are in good agreement with the expected value of the 
DB energy 
threshold 0.27. The dashed line in Fig.\ref{fig4} is the corresponding
fit with parameters $A=0.175$ and $\tilde{E}_{b,th}=0.198$.

In a refined fitting we allow also for breathers with larger
energies. Then the energy per site contribution to $E$ will be
given by
\begin{equation}
A\int_{\tilde{E}_{b,th}}^{\infty} E_b {\rm e}^{-\frac{E_b}{E_0}} 
\rho(E_b) dE_b \;,
\label{fit2}
\end{equation}
where $\rho(E_b)$ is the density of DB states.
Assuming $\rho(E_b)={\rm 1}$ (its actual constant value can
be always absorbed in $A$),
(\ref{fit2}) yields $
A(E_0 E_{b,th}+E_0^2) {\rm e}^{-\frac{\tilde{E}_{b,th}}{E_0}}$
and
the subsequent fitting procedure results similar results
for $\tilde{E}_{b,th}$ 
as obtained with (\ref{fit1}). 
The dashed-dotted line in Fig.\ref{fig4} is the corresponding
fit with parameters $A =5.53$ and $\tilde{E}_{b,th}=0.163$.
While the low energy region of the numerical data
is well reproduced, the high energy data are underestimated
with (\ref{fit1}) and overestimated with (\ref{fit2}).
This implies that the density $\rho(E_b)$ is not constant, but
decaying with increasing energy $E_b$.

\begin{figure}
[!h]
\includegraphics[angle=0, width=1\textwidth]{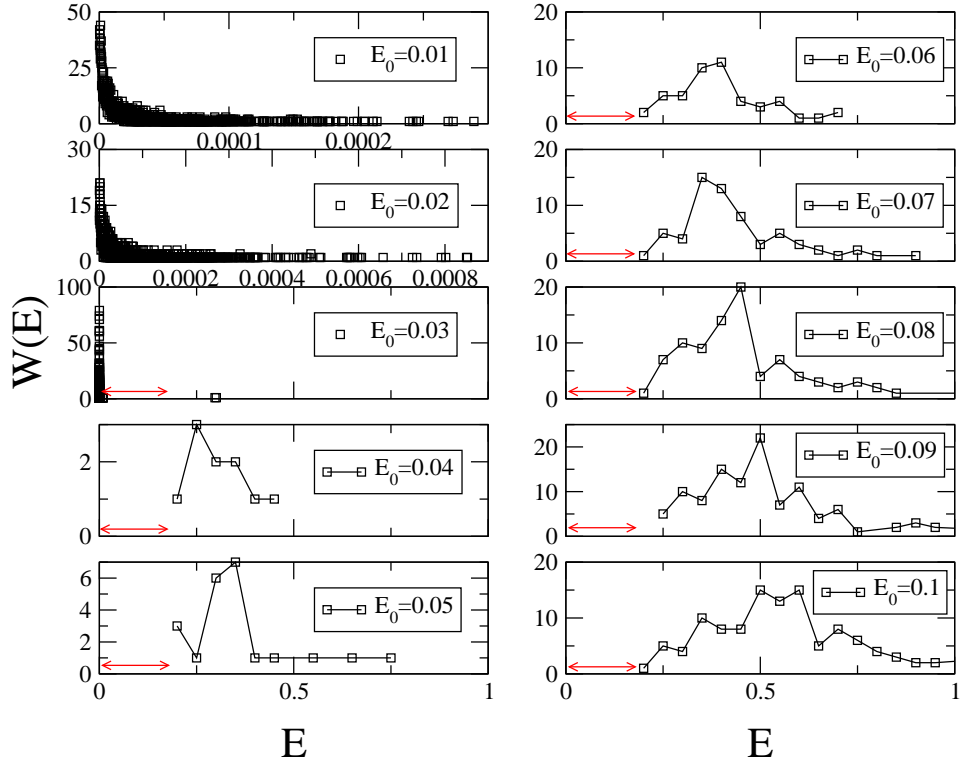}
\caption{Graph of the energy distribution $W(E)$. See text for 
details. The arrows have length 0.17 and 
show the energy region below which the distribution is
expected to vanish due to the DB energy threshold. 
}
\label{fig5}
\end{figure}

\begin{figure}
\includegraphics[angle=0, width=1\textwidth]{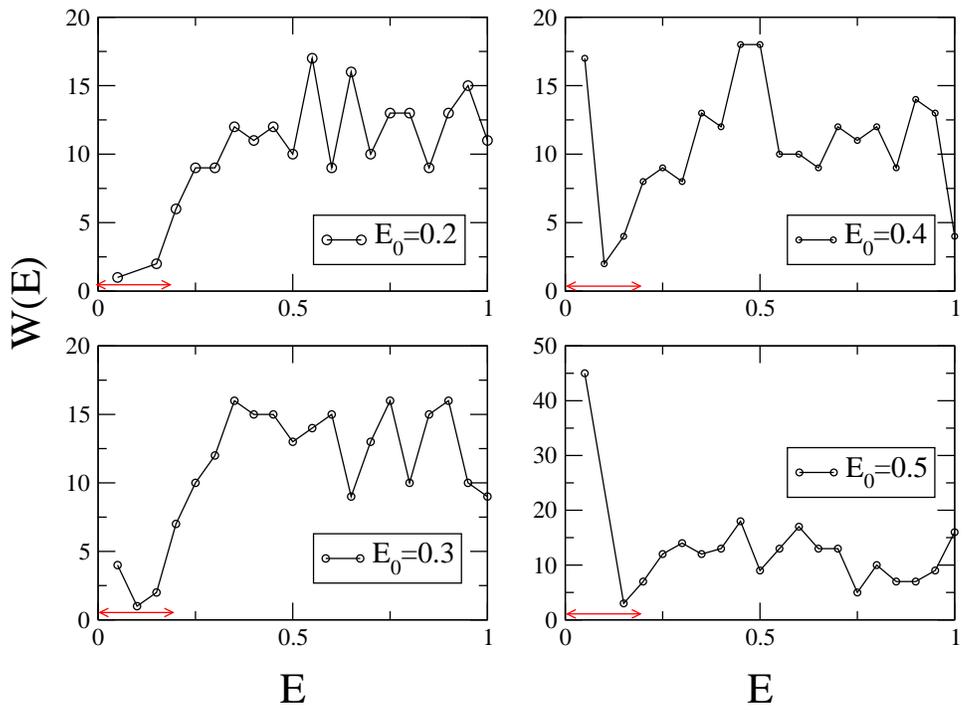}
\caption{Graph of the energy distribution $W(E)$ as in Fig. \ref{fig5} but 
for higher initial energy $E_{0}$. }
\label{fig6}
\end{figure}

But we can obtain even more relevant data.
After the cooling process we measure all values $h_{l,m}$
of the energy density.
In Figs. \ref{fig5} and \ref{fig6} we present their distribution $W(E)$ 
additionally averaged over all the realizations after cooling. 
The initial energy per site that 
is used is indicated in the labels in each subfigure. 
Note that the data are obtained by coarsegraining the energy
axis with a grid size of 0.05. The data obtained within the
first box $0 < E < 0.05$ are omitted from the plots (except for
the two left upper panels at the two lowest energies $E_0$).

The 
DB energy threshold is 
visible in all the subgraphs except the first two subgraphs where the initial 
energy $E_{0}$ is very low and DBs cannot be formed. 
We also clearly see that already for energies $E_0=0.04$ a whole distribution
of breather energies is obtained, with maximum breather energies
being twice larger compared to the minimum DB energy.
Note that the observed energy gaps are very close to the
value 0.17. This corresponds exactly to the expected energy
of the central site for the minimum energy breather being equal
to 0.167 (see Fig.\ref{figbr3d}B). The nearest neighbours of such
a breather carry approximately ten percent of the full DB energy each,
i.e. 0.027. Their contributions are thus within the mentioned first
box and not present in the plotted data.

\begin{figure}
\includegraphics[angle=0, width=1\textwidth]{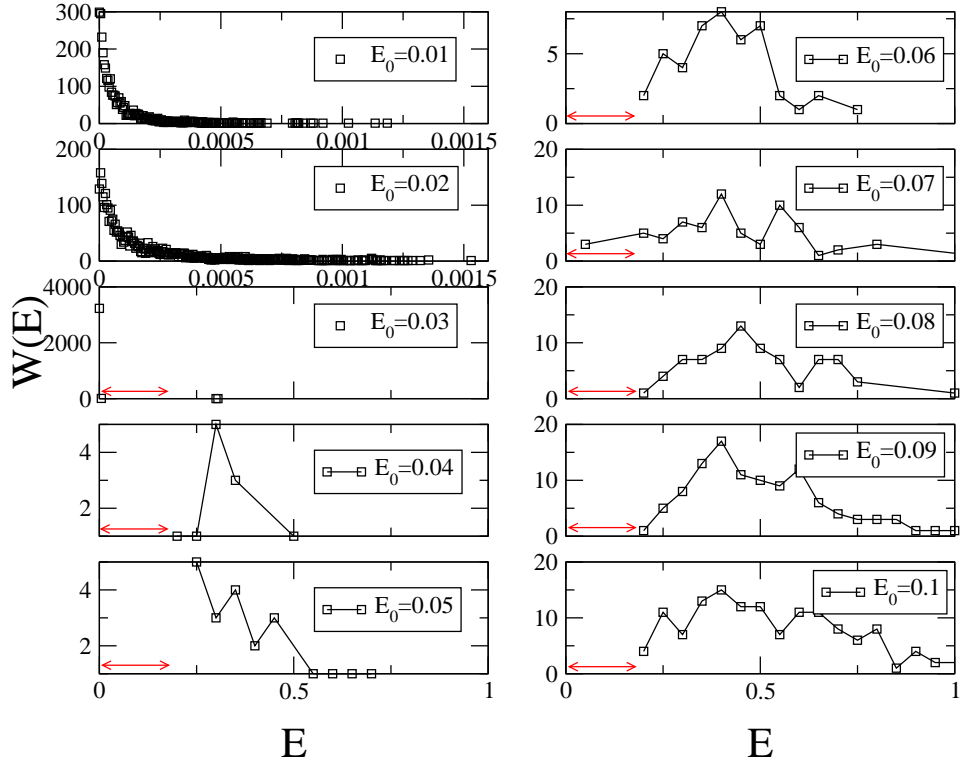}
\caption{Graph of the energy distribution $W(E)$ as in Fig. \ref{fig5} but 
for $T_{cool}=5000$. }
\label{fig5b}
\end{figure}

\begin{figure}
\includegraphics[angle=0, width=1\textwidth]{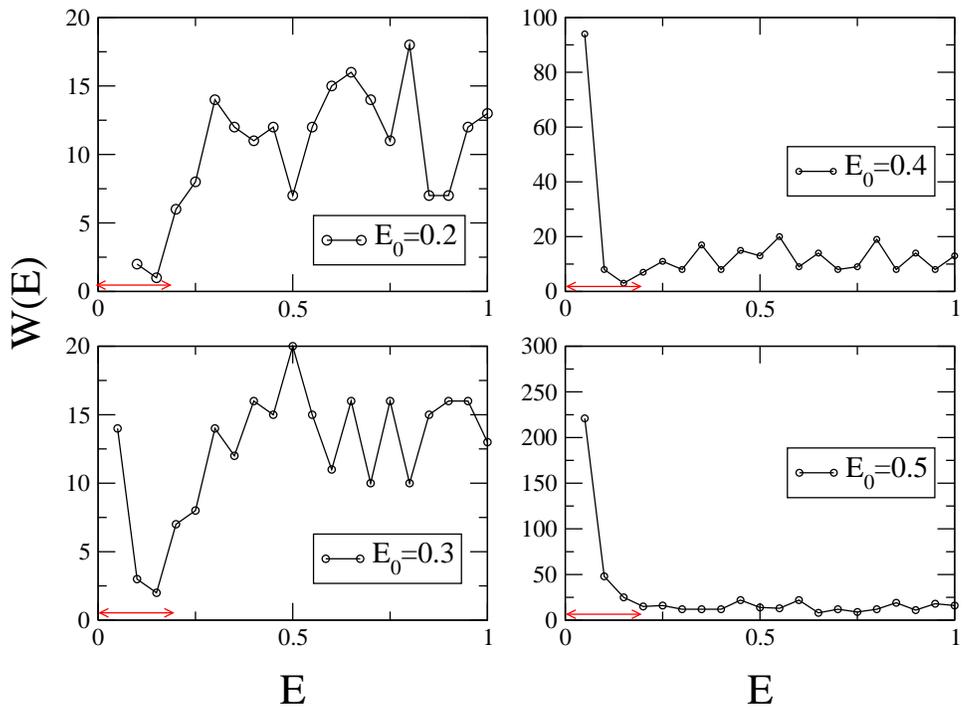}
\caption{Graph of the energy distribution $W(E)$ as in Fig.\ref{fig6} but 
for $T_{cool}=5000$. }
\label{fig6b}
\end{figure}

In order to test the sensitivity of these data on the 
cooling time $T_{cool}$ we present in Figs.\ref{fig5b},\ref{fig6b}
similar results for $T_{cool}=5000$ which is twice smaller.
While the overall statistics needs to be improved in both cases,
we find semiquantitative agreement. This indicates that during
the rather long cooling times the statistical properties of
the localized excitations remaining in the lattice are not
significantly changing.
The observed increase of $W(E)$ for small $E$ and large 
temperatures may be due to the fact that at these large
temperatures the cooling time $T_{cool}$ was too short,
leaving delocalized excitations inside the system.
We will provide with further evidence for the correctness
of this conclusion.

\begin{figure}
\includegraphics[angle=0, width=1\textwidth]{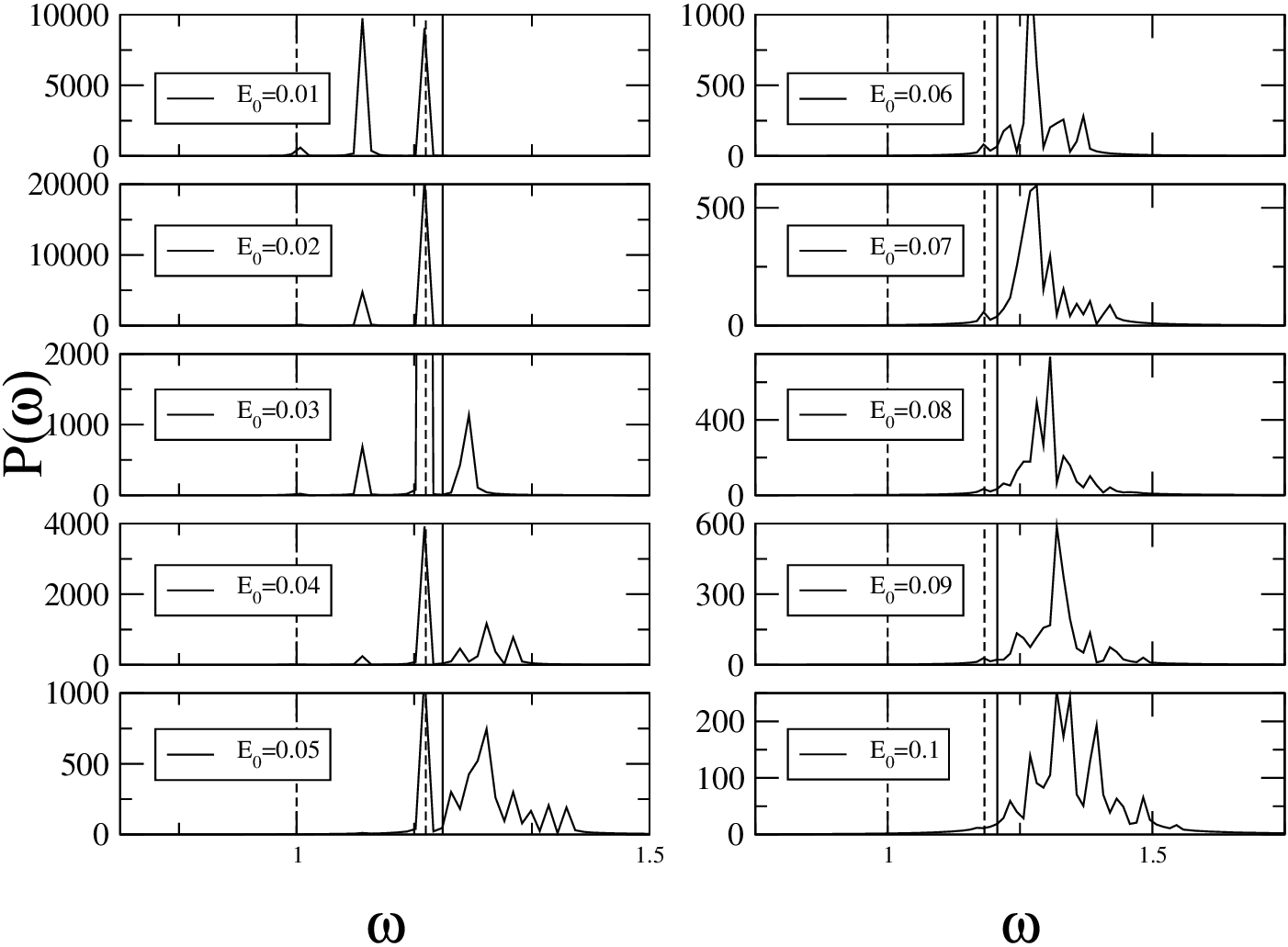}
\caption{Same as in Fig.\ref{fig1} but after 
$T_{cool}=8000$. }
\label{figbrpowera}
\end{figure}

\begin{figure}
\includegraphics[angle=0, width=1\textwidth]{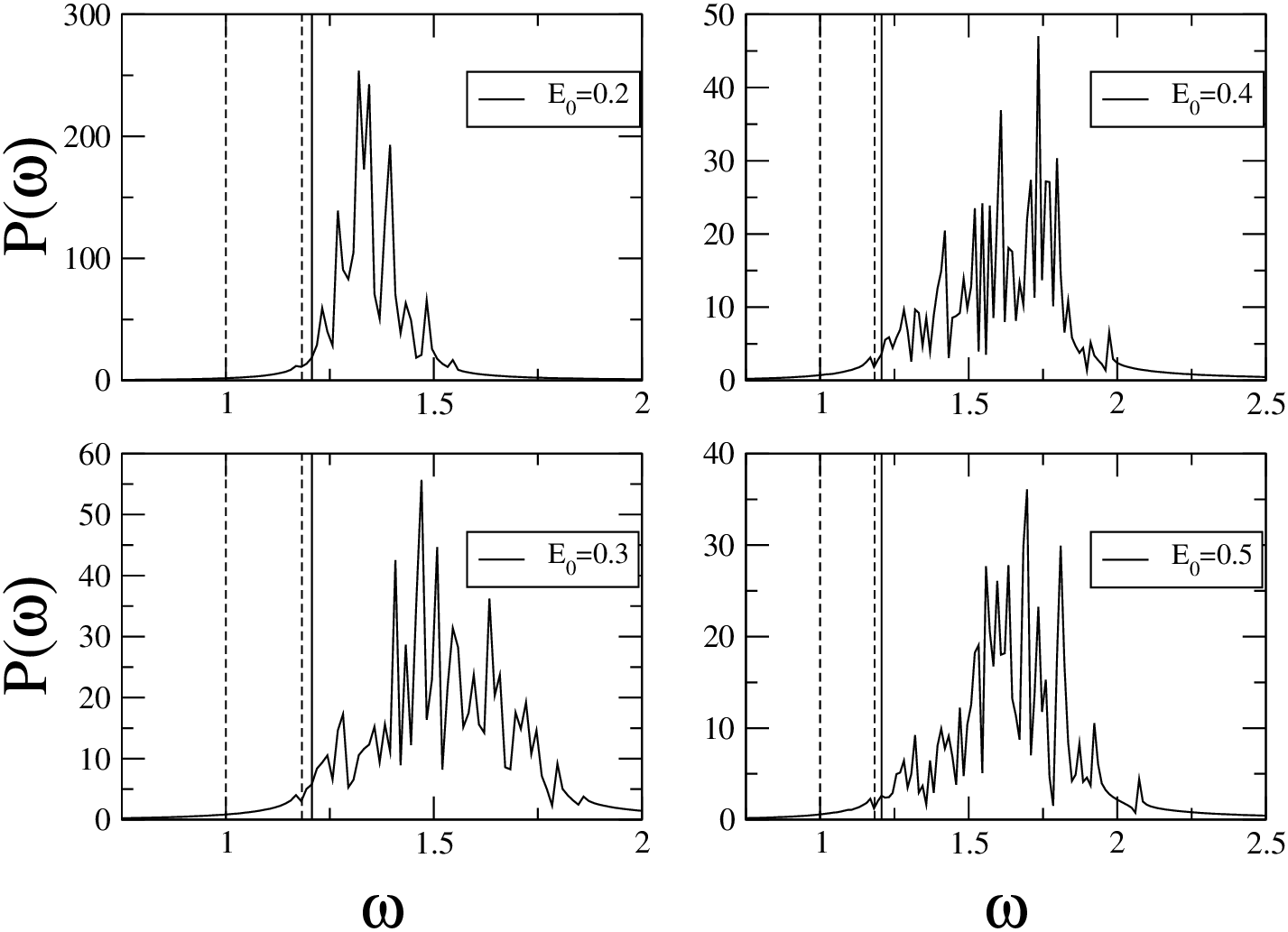}
\caption{Same as in Fig.\ref{fig1} but after 
$T_{cool}=8000$. }
\label{figbrpowerb}
\end{figure}

The above analysis suggested that to some accuracy the excitations
in the lattice at thermal equilibrium can be considered as
a sum of localized and delocalized excitations. The power spectra
in Fig.\ref{fig1} represent thus a sum of the power spectra
of both types of excitations. We use now the cooling process,
which leaves us with the localized excitations only. We wait
for $T_{cool}=8000$ and after that compute the power spectrum
of the remaining localized excitations in the system.
The results are shown in Figs.\ref{figbrpowera},\ref{figbrpowerb}.
The comparison of Figs.\ref{figbrpowera},\ref{figbrpowerb} with
Fig.\ref{fig1} shows, that for energies $E_0 < 0.03$
i) the dynamics is essentially governed by delocalized excitations;
ii) after the cooling period only delocalized states with
almost zero group velocities remain in the system (these correspond
to the band edge frequencies and to the peak inside the band).
For larger energies $E_0$ the power spectra after cooling show
the existence of localized excitations with frequencies outside
the band $\omega_{\vec{q}}^2$. Moerover, we observe very clearly
that the frequency gap $\omega_{\pi} < \Omega_b < \Omega_{b,th}$
is depleted here, indicating that unstable DBs decay during
the cooling time and radiate their energy into the dissipative boundary,
even if they were present to some extent in thermal equilibrium.
Thus the presented way of separating localized
from delocalized excitations in thermal equilibrium
allows also to quantitatively determine the contribution of DBs to
correlation functions. Finally we observe that at the largest
temperatures considered, some nonzero statistical weight is
observed in the frequency region of the phonon band and even
below it. This confirms our previous expectation, that at these
large temperatures the cooling times are probably not long enough
to let the extended excitations out of the system.
A possible reason may be that at these large temperatures many
DBs are coexisting in the system. Extended waves will spend more
time to diffuse around these scattering centers. DBs could
even form temporary percolation networks which 
would hinder the escape of waves even more efficiently.

\section{Conclusions}

Our results show that discrete breathers (DBs) leave clear and
detectable fingerprints in the thermal equilibrium properties
of nonlinear lattices. Using the case of a two-dimensional
lattice, we demonstrated the persistence of an energy threshold
for the existence of DBs in time-dependent correlation functions
by observing weak crossover features.
Moreover, we used the technique of boundary cooling to separate
the localized excitation part at thermal equilibrium from
the delocalized ones. This allows us to study statistical
properties of DB excitations in thermal equilibrium, e.g. their
contribution to the abovementioned correlation functions.
This simple step allowed us to unambiguously confirm the presence
of the energy threshold for DBs in thermal equilibrium.
We furthermore confirm that the frequency gap (which 
corresponds to the excitation of unstable DBs) is depleted
after the separation procedure. Thus we conclude that unstable
DBs are not contributing to dynamical correlations at thermal
equilibrium. By comparing the spectra of the system at thermal
equilibrium with the spectra of the DB part only, we can
provide with reliable statistical weights of both the delocalized
and localized excitation parts of the system at thermal equilibrium
at one and the same frequency. It seems to be very difficult
to provide with a similar cooling and separation 
procedure in any realistic experimental
setup, despite the fact that DBs have been observed in a variety
of different systems. That makes computational studies
of DB properties at thermal equilibrium a unique way of
gaining further understanding of the relaxation and excitation
properties of complex lattices.

With this simple technique, which certainly needs more
refinement, we are now able to reliably compute various
statistical contributions of DBs, including also correlation
effects between DBs and delocalized excitations.
The door is thus open for starting serious analytical work
on DBs in thermal equilibrium, since the presented new numerical
tools will allow for a much more refined testing of various theories
as compared to simulations of thermal equilibrium only.
\\
\\
\\
Acknowledgements
\\
We thank V. Fleurov, G. Kalosakas,
R. Livi and G. Tsironis for useful discussions.

\end{document}